\documentclass[aps,prx,superscriptaddress,reprint,longbibliography]{revtex4-2}
\usepackage{framed}

\usepackage{graphicx}
\usepackage{dcolumn}
\usepackage{bm, bbm}
\usepackage{amsmath, amsfonts, amssymb}
\usepackage{subfigure}
\usepackage{color, xcolor}
\usepackage{mathdots}
\usepackage{mathtools}
\usepackage{braket}
\usepackage{amsthm}
\usepackage{hyperref}

\newcommand{\tr}{\operatorname{tr}}

\newcommand{\id}{\mathbbm{1}}
\newcommand{\Exp}{\mathop{\mathbb{E}}}

\renewcommand{\Pr}{\mathrm{Pr}}

\graphicspath{{./figures/}}

\begin{document}

\title{Measuring unconventional causal structures in monitored dynamics}

\author{Hong-Yi Wang}
\affiliation{Princeton Quantum Initiative, Princeton University, Princeton, NJ 08544, USA}
\author{Haifeng Tang}
\affiliation{
Leinweber Institute for Theoretical Physics, Stanford University, Stanford, California 94305, USA}
\author{Xiao-Liang Qi}
\email{xlqi@stanford.edu}
\affiliation{
Leinweber Institute for Theoretical Physics, Stanford University, Stanford, California 94305, USA}

\date{\today}

\begin{abstract}
Causality underpins all logical reasoning. However, the causal structure in quantum processes can be far from intuitive, often differing from its classical counterpart in relativity, which is defined by the light cone. 
In particular, in systems with measurement and post-selection, causal influence can occur between spacelike separated regions.  
In this work, we study the causal structure and emergent ``arrow of time'' in monitored quantum dynamics, particularly their dependence on initial and final states. We propose a new measure, the cross-entropy quantum causal influence, to quantify the extent of causal influence, whose simulation demonstrates exotic causal structures, such as inverted light cones. This quantity can be measured in current quantum computing platforms. Additionally, we provide an analytical understanding of the relation between time arrow and entropy by studying two types of models that are analytically tractable: a quantum Brownian evolution model and a dual-unitary circuit model. 

\end{abstract}

\maketitle


\section{Introduction}
\label{sec:intro}

Causality is a fundamental aspect of physical spacetime and underpins all logical reasoning. Relativity provides a unifying framework for causality, namely that a local perturbation can influence only a future-directed conic region bounded by lightlike geodesics. In lattice models with a local Hamiltonian, there is no sharp light cone; instead, an approximate light cone emerges, outside of which the influence decays exponentially. This behavior is formalized by the Lieb–Robinson bound \cite{Lieb1972}. 

\begin{figure}[t]
    \centering
    \includegraphics[width=\linewidth]{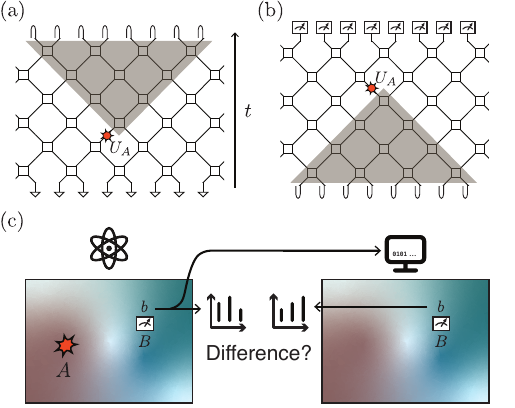}
    \caption{
    (a) In a standard brickwork circuit evolution, the conventional light cone is formed by the spacetime coordinates that a local unitary perturbation $U_A$ can influence. Here, each square represents a two-qubit gate, each triangle represents a single-qubit pure state, and the bending curve at the final time denotes that the circuit is contracted with an identical (but Hermitian-conjugated) backward evolution circuit. 
    (b) For the same brickwork circuit, if the initial state is maximally mixed and the final state is post-selected, the region that $U_A$ can influence forms a light cone propagating backward in time.
    (c) Schematic of the central idea underlying our proposed cross-entropy measure of QCI. A physical system subjected to a local unitary perturbation $U_A$ is compared with an otherwise identical simulated copy in which $U_A$ is not applied. We perform measurement in another region $B$. The QCI from $A$ to $B$ is quantified by how ``different'' the measurement outcome (``$b$'') distributions are. 
    }
    \label{fig:mainidea}
\end{figure}

It is natural to ask whether the causal structure in quantum many-body systems can be qualitatively distinct from that in classical systems. Motivated by this question, Ref.~\cite{Cotler_2019} defined an operational measure of quantum causal influence (QCI). The key idea is to define the influence of a spacetime region $A$ on another region $B$ by applying a unitary operator in $A$ and measuring its impact on the expectation values of operators in $B$. In contrast to classical systems, where causal influence is determined by the light cone, quantum many-body systems can exhibit various unconventional QCI phenomena. Two prominent features of QCI are nonlocality and state dependence. Nonlocality means that unitary operations on a region $A$ can have a trivial impact on each local region $B_1, B_2, \ldots, B_n$, while having a nontrivial impact on a nonlocal operator supported on $B_1 \cup B_2 \cup \ldots \cup B_n$. State dependence means that the same dynamics, with different initial and final states (and more generally, different spatial boundary conditions), can exhibit different QCI.

To illustrate the significance of the state dependence of QCI, we begin with a simple example of unconventional causal structure shown in Fig.~\ref{fig:mainidea}(a,b). Both systems undergo a brickwork quantum circuit evolution but differ in their initial and final states. For a pure initial state, an operator insertion in $A$ influences a future light cone, as expected (Fig.~\ref{fig:mainidea}(a)). However, in Fig.~\ref{fig:mainidea}(b), where the initial state is maximally mixed and the final state is post-selected to a specific pure state (note that the measurement outcome is broadcast to the entire system, as in quantum teleportation), one expects the causal future of $A$ to form a backward light cone in time, because the contours in (a) and (b) are equivalent up to a reversal of time. This observation shows that final states can influence the causal structure in a highly nontrivial way. Although endowing quantum evolution with a post-selected final state may seem contrived at first sight, it has been extensively studied in recent years in the context of measurement-induced phase transitions and monitored quantum dynamics \cite{li2018quantum,skinner2019measurement,fisher2023random}. It is also related to discussions in black hole physics and quantum gravity, such as the black hole final-state proposal \cite{horowitz2004black}. Emergent causal structure is an essential ingredient of emergent spacetime in holographic duality. The relation between the bulk speed of light and boundary quantum chaos has been studied using holographic tensor networks \cite{roberts2016lieb,mezei2017entanglement,qi2017butterfly}.


In this paper, we provide a new measure of QCI in general monitored dynamics, together with an experimental protocol that is realizable on today's noisy quantum devices. One of the main obstacles in monitored-dynamics experiments is the post-selection sampling overhead. To address this issue, we propose the concept of \emph{cross-entropy quantum causal influence (XEQCI)}. As shown in Fig.~\ref{fig:mainidea}(c), causal influence can be quantified by the difference between outcome distributions with and without a perturbation. In XEQCI, we classically simulate the latter case, which effectively renders QCI linear in the quantum register and thereby avoids the post-selection problem. The insight underlying XEQCI is similar to that of cross-entropy benchmarking of quantum devices, which was proposed and experimentally realized in \cite{Boixo2018,Arute2019}. Related approaches using cross-entropy to mitigate post-selection overhead have also been explored in \cite{PhysRevLett.125.070606,Hoke2023,PhysRevLett.130.220404,PRXQuantum.5.030311}. Similarly, our proposed measurement protocol is realizable on current quantum hardware, in particular superconducting qubits and neutral-atom platforms.

We numerically simulate the XEQCI protocol using Clifford circuits to gain intuition for how monitoring affects causal structure. The key findings can be summarized as follows:
(1) the ``arrow of time'' generically points from low-entropy regions to high-entropy regions; and
(2) projective measurements can ``reflect'' the propagation of causal influence.
We carry out numerical simulations on two classes of systems. In the first class, the time evolution is unitary except for a projective measurement at the final time. In this case, we observe that the light cone can undergo a continuous deformation from forward to inverted (i.e., pointing backward in time), depending on the initial and final states. In the second class, we study hybrid quantum circuits composed of unitary gates and measurements. Such systems are known to exhibit a measurement-induced phase transition (MIPT) as the measurement rate $p$ varies. We find that QCI is screened in both phases and can propagate over long distances only when $p$ is near the critical point. The screening mechanisms in the two phases have different physical interpretations: in the volume-law phase, causal influence decays due to quantum information scrambling, whereas in the area-law phase it decays due to measurement. Therefore, QCI provides a new lens for studying MIPT. Near the critical point $p_c$, the dynamical exponents of the QCI transition are found to coincide with those of the measurement-induced transition in mutual information \cite{PhysRevB.100.134306}.

To obtain a deeper understanding of the behavior of QCI in different systems, we study two additional analytically tractable models. First, we consider Brownian Gaussian unitary dynamics, in which we quantitatively confirm that the averaged QCI is proportional to the purity difference between the initial and final states. As a result, the direction of the time arrow in this system is explicitly determined by the second R\'enyi entropy. We also study QCI in dual-unitary circuits. By introducing a monitored spacetime region in such circuits, we show that the monitored region can act as a ``source'' of the arrow of time, meaning that the future light cones for small regions outside the monitored region point outward, away from the monitored region. This illustrates that causal influence can exhibit richer behavior than simply selecting between future and past directions.

The remainder of this paper is organized as follows. In Sec.~\ref{sec:framework}, we review the concept of QCI and define the new cross-entropy measure XEQCI. Sec.~\ref{sec:experiment} describes the proposed experimental protocol for efficiently measuring XEQCI. Sec.~\ref{sec:simulation} presents numerical results from Clifford circuit simulations of the proposed experiment. This section is divided into two parts: Sec.~\ref{ssec:light_cones} studies a unitary circuit with post-selection at the final time, while Sec.~\ref{ssec:MIPT} focuses on QCI in hybrid circuits exhibiting MIPT. Sec.~\ref{sec:analytical} studies two classes of models in which QCI is analytically tractable, including a Brownian evolution model with initial and final states (Sec.~\ref{ssec:brownian}) and a dual-unitary circuit with measurements in a spacetime region (Sec.~\ref{ssec:dual}). Finally, Sec.~\ref{sec:conclusion} contains the conclusions and further discussion.

\section{Framework of QCI}
\label{sec:framework}

We begin with a general discussion of how to quantify causal influence in quantum systems. The most general question to ask is whether a set of perturbations affects another set of measurements. Since we are primarily concerned with whether a spacetime region $A$ influences another region $B$, we take the perturbations to be all possible unitary operators supported in $A$, and the measurements to be all allowed quantum measurements in $B$.

This paves the way for the following mathematical description. Consider a fixed, general quantum evolution (which is not necessarily unitary). In addition to the original evolution, we apply an additional unitary operator $U_A$ in region $A$ and a positive operator-valued measure (POVM) in region $B$. The Born probabilities of measurement outcomes can be calculated within the Schwinger-Keldysh formalism with appropriate normalization. Suppose the POVM is given by a set of Kraus operators $\{K_b\}_{b=1}^{N_B}$.
Each $K_b\in\mathcal{B}(\mathcal{H}_B)$, where $\mathcal{H}_B$ denotes the Hilbert space in $B$, and $\mathcal{B}$ denotes the set of bounded operators. The Born probability can be represented by the following contour:
\begin{equation}    \label{eq:born_keldysh}
    \Pr(b|U_A) = 
    \vcenter{\hbox{\includegraphics{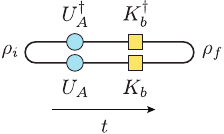}}},
\end{equation}
where the straight lines represent time evolution operators, and the bends at the left and right ends represent the initial state and the final state (due to possible post-selection; if there is no post-selection then $\rho_f=\id$), respectively. Alternatively, we can view Eq.~\eqref{eq:born_keldysh} as definig a bilinear function $\varrho_B$ in $K_b$ and $K_b^\dagger$, which is normalized as $\sum_{b=1}^{N_B} \varrho_B (K_b, K_b^\dagger) = 1$. 
The object $\varrho_B$ has previously been referred to as the \emph{superdensity operator} in $B$ \cite{Cotler_superdensity}. Whether $U_A$ has any effect on the measurement outcomes in $B$ can be determined by comparing $\varrho_B(U_A)$ with $\varrho_B(U'_A)$. If $\varrho_B(U_A)$ does not depend on $U_A$, no experiments performed in $B$ are affected by any perturbation in $A$. Hence, any distance measure between density operators, such as the trace distance, can be used to define a viable measure of QCI.

Ref.~\cite{Cotler_2019} introduces two quantities to quantify QCI, which both fall into the framework we described above. Specifically, Ref.~\cite{Cotler_2019} proposes the maximal QCI
\begin{equation}    \label{eq:cotler_max}
    \begin{aligned}
    \mathrm{CI}(A:B)
    &= \sup_{U_A, O_B} \frac{1}{\left\| O_B \right\|_2^2} \\
    & \left| \varrho(U_A)(O_B, O_B^\dagger) - \int d{U_A} \, \varrho(U_A)(O_B, O_B^\dagger) \right|
    \end{aligned}
\end{equation}
and the averaged QCI
\begin{equation}    \label{eq:cotler_avg}
    \begin{aligned}
    \overline{\mathrm{CI}}(A:B) &= \int dU_A \int dO_B \\
    &\left| \varrho(U_A)(O_B, O_B^\dagger) - \int d{U_A} \, \varrho(U_A)(O_B, O_B^\dagger) \right|^2
    \end{aligned}
\end{equation}
where $U_A$ is integrated over the Haar measure of $\mathrm{U}(\mathcal{H}_A)$, and $O_B$ is integrated uniformly over the sphere of $\left\| O_B \right\|_2 = 1$. In some random dynamics, the averaged QCI can be analytically calculated using random matrix theory. 

However, neither of these quantities is directly measurable experimentally. Hence, we propose a new cross-entropy--like quantity to characterize QCI: 
\begin{equation}    \label{eq:chi_def}
    \chi_{U_A} = \frac{\sum_b \Pr(b|U_A) \Pr(b|\id_A)}{\sum_b \Pr(b|\id_A)^2}    .
\end{equation}
This quantity measures the similarity between the two vectors $\{\Pr(b|U_A)\}_{b=1}^{N_B}$ and $\{\Pr(b|\id_A)\}_{b=1}^{N_B}$ by calculating their inner product. $\chi_{U_A}=1$ when $\Pr(b|U_A)$ and $\Pr(b|\id_A)$ are identical, while $\chi_{U_A}$ is close to zero when the two probability distributions are significantly different. In the next section, we construct an explicit experimental protocol to measure an average of the quantity in Eq.~\eqref{eq:chi_def} for generic monitored dynamics. We remark that a similar cross-entropy to study the decodability transition in MIPT was proposed in Ref.~\cite{PhysRevLett.130.220404} and realized in Ref.~\cite{PhysRevLett.134.120401}.

\section{Experiment proposal}
\label{sec:experiment}

In this section, we introduce an experimental protocol to measure QCI based on $\chi_{U_A}$ defined in Eq. (\ref{eq:chi_def}). 

To translate the general idea into an experimental protocol, we need to overcome two key obstacles. First, we note that any measure of QCI is necessarily a non-linear observable (in terms of $\varrho_B$, i.e., the Keldysh contour of the evolution). Hence, we need to find ways to correlate multiple copies of the system. Second, when the evolution is non-unitary, the dynamics must be experimentally implemented via unitary evolution plus post-selection on some designated measurement outcomes. This thus leads to a substantial post-selection sampling overhead. 

Our proposal avoids this overhead by calculating quantum--classical correlations. To make this concrete, suppose the non-unitary evolution consists of unitary evolution interleaved with post-selection, the outcome of which is labeled by $m$. Recall that we always use $b$ to label the measurement outcome of the POVM in region $B$, which is meant to detect QCI, whereas $m$ is the outcome of other measurements that define a trajectory of the non-unitary evolution. Although the distribution of $m$ could also be influenced by $U_A$, our goal is to study \emph{the influence on $B$ within a single trajectory labeled by $m$}. Each instance of quantum experiment samples $(b,m)$ with joint probability distribution $\Pr(b,m|U_A)$. When we get $(b,m)$, we classically calculate the quantity
\begin{equation}    \label{eq:csim}
    c_{b,m} = \frac{\Pr(b|m,\id_A)}{\sum_b \Pr(b|m, \id_A)^2}   . 
\end{equation}
We then measure $b$ and $m$ in the experimental system for many realizations of $U_A$ to estimate the average of $c_{b,m}$. This average converges to 
\begin{equation}    \label{eq:chi}
    \boxed{ \chi = \int dU_A \, \sum_{b,m} \Pr(b,m,U_A) c_{b,m} . }
\end{equation}
We refer to Eq.~\eqref{eq:chi} as the \emph{XEQCI}, the central quantity in our experimental proposal and the following numerical studies. The probability distribution over the perturbation $U_A$ can depend on the experimental setup. To obtain an unbiased measure of QCI between these regions, we prefer the ensemble of $U_A$ to be uniform. In the Clifford circuit simulations in Sec.~\ref{sec:simulation}, we will choose $U_A$ from the ensemble of random Clifford operators. 
More details will be discussed in Sec.~\ref{sec:simulation}. The POVM applied in $B$ is also preferably chosen to enumerate over all possible Kraus operators. 

We now explain why the above quantity $\chi$ is associated with QCI and is in fact an average of the cross entropy Eq.~\eqref{eq:chi_def}. Within one definite trajectory $m$, Eq.~\eqref{eq:chi_def} can be written as
\begin{equation}
    \chi_{m,U_A} = \frac{\sum_b \Pr(b|m, U_A) \Pr(b|m, \id_A)}{\sum_b \Pr(b|m, \id_A)^2}  .
\end{equation}
Using the chain rule of conditional probability, Eq. (\ref{eq:chi}) becomes
\begin{align}\label{eq:chain_rule}
    \chi &= \int dU_A \, \sum_{b,m} \Pr(b,m,U_A) \frac{\Pr(b|m,\id_A)}{\sum_b \Pr(b|m, \id_A)^2} \nonumber\\
    &= \Exp_{U_A} \sum_m \Pr(m|U_A) \frac{\sum_b \Pr(b|m, U_A) \Pr(b|m, \id_A)}{\sum_b \Pr(b|m, \id_A)^2}  \nonumber\\
    &= \Exp_{U_A} \sum_m \Pr(m|U_A) \chi_{m,U_A}    .
\end{align}
As claimed, $\chi$ is the average of the cross entropy $\chi_{m,U_A}$ over $U_A$ and $m$ according to the Born probability of each trajectory. 

Finally, the XEQCI in Eq.~\eqref{eq:chi} can be averaged over a class of circuits as well, since in many cases we are interested in the causal structure of a class of quantum evolution rather than a single circuit. We summarize the experimental protocol to measure the circuit-averaged XEQCI, i.e., $\overline{\chi} = \Exp_{C}\chi_C$, as follows. 
\begin{framed}
\begin{enumerate}
    \item Sample a quantum circuit $C$ from a distribution $\mathcal{C}$, which may contain measurements to decide trajectories of non-unitary evolution. 
    \item Sample $U_A$, the perturbation in region $A$. 
    \item Implement the circuit and apply $U_A$ in region $A$. Measure in region $B$ to obtain outcome $b$ and elsewhere to obtain outcome $m$. 
    \item Calculate $c_{b,m}$ according to Eq.~\eqref{eq:csim}.
    \item Repeat from step 1 as needed. Finally, average over all obtained $c_{b,m}$. 
\end{enumerate}
\end{framed}

\section{Simulation in Clifford Circuits}
\label{sec:simulation}

As discussed in the previous section, our measurement protocol for QCI is effective whenever the quantum dynamics can be classically simulated. Here, we present numerical simulations of this protocol using the stabilizer formalism, which can be accomplished in polynomial time \cite{gottesmanknill1998}. The unitary evolution is a brickwork circuit of two-qubit Clifford gates. These stabilizer simulation results suggest that they capture general features of QCI under unitary dynamics and measurements. We study two distinct scenarios in the following subsections. Sec.~\ref{ssec:light_cones} focuses on unitary circuits with post-selection only at the final time, whereas Sec.~\ref{ssec:MIPT} focuses on circuits with mid-circuit measurements.

We visualize most of our data as a ``QCI landscape''. In a QCI landscape plot (see, e.g., Fig.~\ref{fig:light_cones}), the information is displayed as follows. We apply a perturbation $U_A$ in a region $A$, outlined in cyan. The region $B$ is chosen to be a single site and is scanned over all spacetime coordinates $(x_B, t_B)$. We then construct a color plot in which the color at $(x_B, t_B)$ represents the value of $-\log_2 \overline{\chi}$ when $B$ is located at that coordinate. Here $\overline{\chi}$ denotes $\chi_{m,U_A}$ averaged over all $U_A$, all measurement outcomes $m$, and a class of circuits $C \sim \mathcal{C}$:
\begin{equation}    \label{eq:chibar}
    \overline{\chi} = \Exp_{C\sim\mathcal{C}} \chi_C   ,
\end{equation}
where $\chi_C$ is defined as in Eq.~\eqref{eq:chi}. Details of how to perform the classical simulation of $\overline{\chi}$ are provided in Appendix~\ref{app:stabilizer}. Intuitively, coordinates with nonzero values in a QCI landscape plot represent the \emph{causal future} of region $A$. For example, in the conventional case of a random brickwork circuit with a pure initial state and no post-selection, the QCI landscape exhibits nonzero values only within a conic region originating from $(x_A, t_A)$ (see Fig.~\ref{fig:light_cones}(a)), which is known as the future light cone.

\subsection{Light cones}
\label{ssec:light_cones}

\begin{figure*}[t]
    \centering
    \includegraphics[]{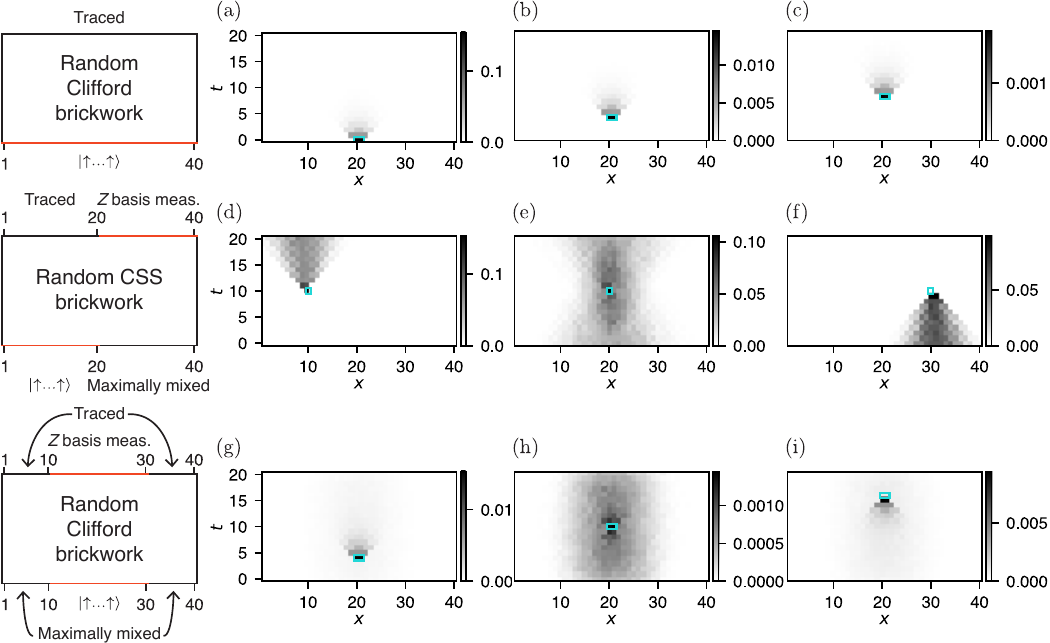}
    \caption{QCI landscapes and light cone inversion. In each labeled figure, the color is the value of $-\log\overline{\chi}$, where $\chi$ is the circuit-averaged cross-entropy measure of QCI (see Eq.~\eqref{eq:chibar}). }
    \label{fig:light_cones}
\end{figure*}

In this subsection, we study unitary quantum circuits with an initial state and a final measurement. Figure~\ref{fig:light_cones} shows the characteristic causal influence resulting from a local perturbation in region $A$, indicated by a cyan rectangle. These examples clearly demonstrate that both the influenced region (i.e., the causal future of $A$) and the intensity of QCI depend on the ambient state and the evolution around $A$. In particular, a final state, corresponding to post-selection on a specific outcome of a projective measurement at the final time, can induce a nontrivial causal influence beyond the conventional future light cone.

The first row, Fig.~\ref{fig:light_cones}(a--c), displays the QCI landscape for a brickwork circuit composed of random two-qubit Clifford gates. The initial state is a pure product state, and there is no post-selection. All three figures show that the QCI is nonzero only within a conventional light cone. In all cases, the signal decays within a distance of order $O(1)$ from region $A$, which can be understood in terms of quantum information scrambling: after the scrambling time, any local operator $K_B$ is unlikely to detect the effect of $U_A$, as $U_A$ evolves into a highly nonlocal operator. Moreover, the QCI decays approximately exponentially as $t_A$ is shifted to later times. We further observe that the intensity of QCI following a perturbation in $A$ is positively correlated with the purity of the state $\rho_A$ immediately before $U_A$ is applied. This phenomenon is also confirmed by our analytic calculations in a different dynamics, presented in Sec.~\ref{ssec:brownian}.

In the second row, Fig.~\ref{fig:light_cones}(d--f), both the initial and final states are set to a half-pure, half-mixed configuration. To observe long-lasting light cones, the evolution is chosen to be a brickwork circuit of random two-qubit ``CSS gates''. The ``CSS gates'' (see Appendix~\ref{app:CSS} for details) are defined as Clifford gates that map $X$ operators to $X$ operators and $Z$ operators to $Z$ operators. The difference between sampling CSS gates and random Clifford gates is that the former exhibit weaker scrambling than the latter, allowing the QCI to maintain an appreciable value for a longer time. However, we expect that the region that can be influenced is the same in both cases.


To illustrate the role of the initial and final states, we consider a configuration in which the left half of the system is initialized in a product state and post-selected to be maximally mixed, while the right half is prepared and post-selected in the reverse manner. In this case, we observe a crossover from a forward-pointing light cone to a backward-pointing one.
The QCI generally propagates from low-entropy regions to high-entropy regions, as also confirmed by the analytic calculation in Sec.~\ref{ssec:brownian}.
\footnote{Ref.~\cite{yates2025localarrowstimequantum} made a similar discovery by examining the difference between forward and backward QCI in a setup different from ours.}
In (d), the system's local state immediately before $(x_A, t_A)$ is almost pure, while the final state is maximally mixed, so the QCI forms a conventional light cone pointing toward $t > t_A$. In contrast, in (f), the state before $(x_A, t_A)$ is maximally mixed whereas the final state is almost pure, resulting in a reversed light cone extending entirely toward $t < t_A$. In case (e), both the initial and final states near $(x_A, t_A)$ are partially mixed, leading to QCI propagating to both the past and to the future. In an intermediate case such as (e), the ``arrow of time'' is not well defined, as the QCI is non-vanishing throughout the system. 

The third row, Fig.~\ref{fig:light_cones}(g--i), illustrates the state dependence of QCI in a different configuration. Here, the initial and final states are identical, with a pure region in the middle and maximally mixed regions away from the center. In this case, the light cone is predominantly upward at early times (panel~(g)), undergoes a crossover (panel~(h)), and becomes predominantly downward at late times (panel~(i)). This behavior is consistent with the intuition that the arrow of time points in the direction of higher entropy. At early times, the state evolving from the past has lower entropy, whereas the state evolving from the future (obtained by backward time evolution from the final state) is more entangled due to entropy growth. At later times, the situation is reversed.
It should be noted that in (h), the absolute value of $-\log_2 \overline{\chi}$ is much smaller than in (g) and (i) because there the two states from both the past and future have high entropy.

In summary, we have observed two key rules of QCI from our numerical results: 
\begin{enumerate}
    \item The intensity of QCI to the future (resp. past) of $A$ is proportional to the purity of the initial (resp. final) state of $A$. 
    \item Generically, QCI flows from regions of lower entropy to regions of higher entropy.
\end{enumerate}

\subsection{Measurement-induced phase transition}
\label{ssec:MIPT}

\begin{figure*}[t]
    \centering
    \includegraphics[width=\linewidth]{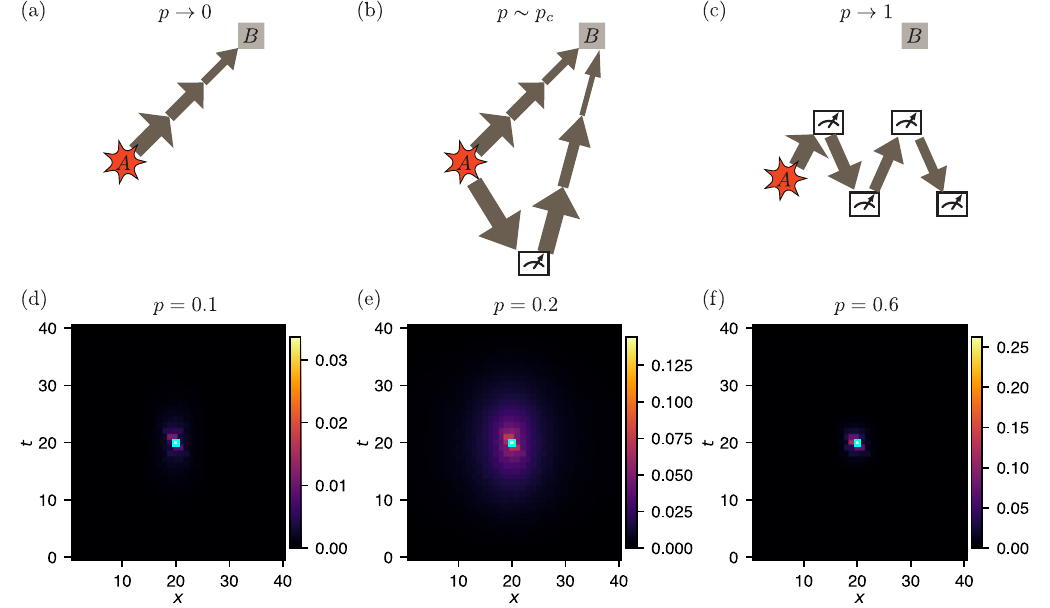}
    \caption{(a-c) Illustration of possible paths through which $A$ can influence $B$, and how they are impacted by mid-circuit measurements. (d-f) The QCI landscape in hybrid circuits with different measurement rates. (a) and (d): small-$p$ phase, where the QCI only spreads an $O(1)$ distance. (b) and (e): close to the critical point $p_c$, where the QCI spreads a maximal distance. (c) and (f): large-$p$ phase, where the spreading is suppressed to an $O(1)$ range again. }
    \label{fig:MIPTidea}
\end{figure*}

In this section, we consider a class of hybrid random circuits composed of a brickwork circuit of random two-qubit Clifford gates, where at each time step and on each qubit a $Z$-basis measurement is applied with probability $p$. It is well known that such circuits exhibit a measurement-induced phase transition (MIPT) as the measurement rate $p$ varies. The low-$p$ and high-$p$ phases are conventionally referred to as the volume-law and area-law phases, respectively, reflecting the scaling of the circuit-averaged entanglement entropy. Below, we study the QCI landscape in these hybrid circuits.

\begin{figure}
    \centering
    \includegraphics[]{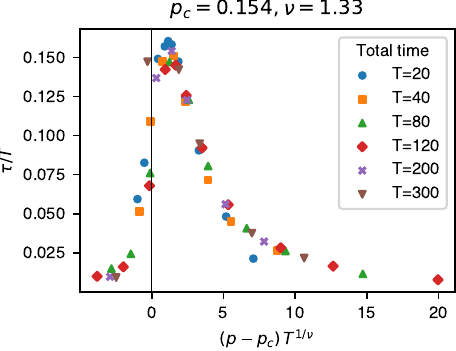}
    \caption{Correlation time of XEQCI and universality. Here, $\tau$ is the correlation time and $T$ is the total time of the simulation. By choosing $p_c = 0.154$ and $\nu = 1.33$, data for different total time $T$ collapse onto one universal curve. 
    }
    \label{fig:correlation_time}
\end{figure}

Let us first examine the numerical results shown in Fig.~\ref{fig:MIPTidea}(d--f). Qualitatively, the QCI propagates isotropically in both space and time and decays with distance. The characteristic decay length of the QCI increases with $p$ at small $p$, but decreases once $p$ exceeds a critical value. Indeed, Fig.~\ref{fig:correlation_time} quantitatively demonstrates, via scaling and universality fitting, the existence of a phase transition point with a diverging characteristic length. The behavior of QCI closely resembles that of the MIPT in mutual information, which also peaks at the critical point $p_c$ \cite{PhysRevB.100.134306}. 

To gain insight into the phase transition in QCI, we introduce a qualitative picture of how QCI behaves under unitary dynamics and measurements. Consider all possible paths connecting regions $A$ and $B$, each of which may contribute to the QCI. As a crude picture, under unitary dynamics, the contribution of a path decays exponentially with its length due to quantum information scrambling. By contrast, the effects of projective mid-circuit measurements are twofold: (1) quantum teleportation---a measurement blocks the direct path but allows reflected paths to contribute to the QCI (see Fig.~\ref{fig:MIPTidea}(b)); and (2) local purification---a local measurement generally reduces scrambling in the local state, thereby slowing the decay of QCI.

Accordingly, we provide the following intuitive picture of the phases and the transition in QCI.
\begin{enumerate}
    \item When there are very few mid-circuit measurements ($p \approx 0$), influence propagation is qualitatively similar to that in a unitary circuit, where the QCI decays over an $O(1)$ range.
    \item When there is a moderate density of mid-circuit measurements ($p \sim p_c$), the measurement-induced reflection of signals enables more paths from $A$ to $B$, and the purification effect slows down the decay. As a result, the range over which QCI remains appreciable becomes much larger.
    \item When there are too many mid-circuit measurements ($p \approx 1$), the QCI is suppressed again. This is because paths are frequently reflected by measurements and are therefore likely to be trapped between measured sites, rendering them unable to reach $B$. In this limit, the characteristic length is essentially the ``mean free path'' that a signal can travel between measured sites, so we anticipate $\tau \propto 1 - p$. In the extreme case $p = 1$, all paths from $A$ immediately terminate at a nearby measured site, and none can reach $B$.
\end{enumerate}

To quantitatively characterize the phase transition in terms of correlation time and length, we consider a slightly modified setup. We impose periodic boundary conditions and set the perturbed region $A$ to be the entire system at time $t_A = T/2$. As a result, the QCI landscape becomes translationally invariant, which clarifies the definition of a characteristic time scale of QCI decay. The correlation time as a function of $p$ is shown in Fig.~\ref{fig:correlation_time} after power-law scaling with the system size $T$ in the time direction. At the critical point, the correlation time diverges and is observed to be approximately proportional to the total time $T$. More details of this computation are provided in Appendix~\ref{app:MIPTfullA}. These results show that the MIPT corresponds to a continuous phase transition in QCI. The universal scaling behavior of the correlation time takes the form $\tau/T = f\left((p - p_c) T^{1/\nu}\right)$ 
\footnote{In general, there is an additional critical exponent $z$ that determines how the correlation time scales with the correlation length. Here, we use the known value $z = 1$ (i.e., the correlation time scales in the same way as the correlation length) for MIPT without independent verification.}.
The phase transition point is found to be $p_c = 0.154$, and the critical exponent is $\nu = 1.33$, both in agreement with Ref.~\cite{PhysRevB.100.134306}. These findings suggest that the divergence of the XEQCI correlation length provides a new perspective on the MIPT.

\section{Analytic results}
\label{sec:analytical}

In this section, we study two types of models where analytic results for QCI can be obtained. 

\subsection{Brownian Gaussian unitary model}
\label{ssec:brownian}
\begin{figure}
    \centering
    \includegraphics[width=1\linewidth]{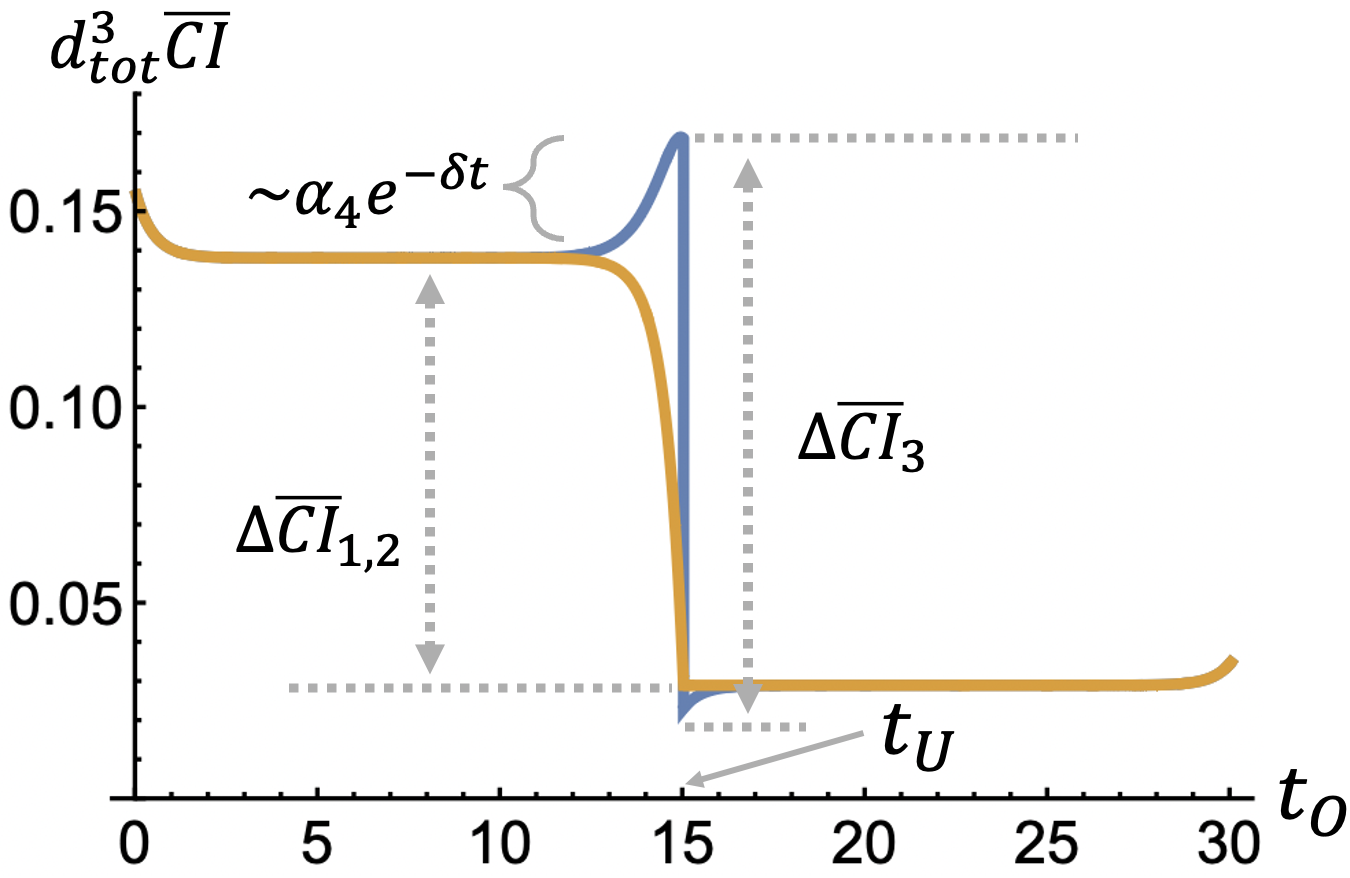}
    \caption{Causal influence in BGUE toy model in~\ref{ssec:brownian}. We fix the whole system being $20$ qubits ($d_\text{tot}=2^{20}$),  total evolution time being $T=30$ and $U_A,O_B$ acting on single qubit ($d_A=d_B=2$). The final state is a product of pure state on each qubit and the initial state is a product mixed state on each qubit with local purity set to a generic value $0.504$. We fix the location of $U_A$ insertion at $t_U=T/2$, and draw the function of $\overline{\text{CI}}$ as a function of $t_O\in[0,T]$. The blue curve is when $U_A,O_B$ acting on the same qubit (so QCI has discontinuity when $t_O$ cross $t_U$) and orange curve is when $U_A,O_B$ acting on different qubits (QCI is continuous when $t_O$ cross $t_U$).}
    \label{fig:BGUE}
\end{figure}

In this subsection, we study a simple analytically tractable model, the Brownian Gaussian unitary ensemble~\cite{Tang2024kpv,Guo2024zmr} (BGUE), to further explain the relation between QCI and the entropy of the initial and final states. As mentioned in Sec.~\ref{sec:framework}, the averaged QCI (Eq.~\eqref{eq:cotler_avg}) turns out to be quadratic in the superdensity operator, and hence can be analytically calculated as long as the second moment is analytically tractable. In this subsection, we study Eq.~\ref{eq:cotler_avg} and expect that it should have qualitatively similar behavior as the XEQCI.

The BGUE is defined on a $d_\text{tot}$-dimensional Hilbert space with a time-dependent Hamiltonian $H(t)$, whose matrix entries are Brownian Gaussian random variables with covariance
$\mathbb{E}[H_{ij}(t) H_{kl}(t')] = d_\text{tot}^{-1} \delta_{il} \delta_{jk} \delta(t - t')$.
This model has no spatial locality and exhibits fast scrambling~\cite{Tang2024kpv,Guo2024zmr}, making it a convenient toy model for capturing key properties of chaotic dynamics.

Now, we compute the averaged QCI defined in Eq.~\eqref{eq:cotler_avg}. Define $\mathcal U_t \equiv \mathcal P e^{-i \int_0^{t} dt' \, H(t')}$ as the time-ordered exponential. Then $\overline{\text{CI}}(A\!:\!B)$ depends only on the second moment of the ensemble of $\mathcal U_t$, which is available in closed analytic form in Ref.~\cite{Tang2024kpv}. Details of the calculation techniques for BGUE can be found in Appendix~\ref{app:BGUE}. Since the BGUE has no spatial locality, we focus on the behavior of $\overline{\text{CI}}(A\!:\!B)$ in the time domain.

Specifically, we consider an initial state $\rho_i$ evolved under BGUE for a time $T$, followed by post-selection onto a final state described by the density matrix $\rho_f$. We apply a Haar-random unitary $U_A$ in region $A$ (of dimension $d_A$) at time $t_U$ ($0 < t_U < T$), and a random Hermitian operator $O_B$ in region $B$ (of dimension $d_B$) at time $t_O \equiv t_U \pm \delta t$ ($\delta t > 0$). We define $\overline{\text{CI}}_{+}$ for $t_U < t_O$ (causal influence toward the coordinate-time future) and $\overline{\text{CI}}_{-}$ for $t_U > t_O$ (causal influence toward the coordinate-time past). We find that the averaged QCI admits a simple analytic form in the following representative parameter regime:
\begin{enumerate}
\item We compare the ``long-term'' QCI difference by setting $\delta t \gg 1$ (with $T \gg t_U, t_O \gg 1$).
\begin{enumerate}
    \item When $U_A$ and $O_B$ probe the same spatial region, we obtain
\begin{equation}
\Delta\overline{\text{CI}}_1 \equiv \overline{\text{CI}}_+ - \overline{\text{CI}}_- 
= \alpha_1 \left[ e^{-S^{(2)}(\rho_i)} - e^{-S^{(2)}(\rho_f)} \right],
\end{equation}
with $\alpha_1 \equiv \frac{(d_A^2 - 1)^2}{d_A^4 (d_A^2 + 1)} \frac{d_\text{tot}}{(d_{\text{tot}}^2 - 1)^2} > 0$.

\item When $U_A$ and $O_B$ probe non-overlapping spatial regions, we obtain
\begin{equation}
\Delta\overline{\text{CI}}_2 \equiv \overline{\text{CI}}_+ - \overline{\text{CI}}_- 
= \alpha_2 \left[ e^{-S^{(2)}(\rho_i)} - e^{-S^{(2)}(\rho_f)} \right],
\end{equation}
with $\alpha_2 \equiv \frac{(d_A^2 - 1)(d_B^2 - 1)}{d_A^2 d_B^2 (d_B^2 + 1)} \frac{d_\text{tot}}{(d_{\text{tot}}^2 - 1)^2} > 0$.
\end{enumerate}

\item We compare the ``instantaneous'' QCI difference by setting $\delta t = 0^+$ (with $T \gg t_U, t_O \gg 1$). When $U_A$ and $O_B$ probe the same spatial region, we obtain
\begin{equation}
\Delta\overline{\text{CI}}_3 \equiv \overline{\text{CI}}_+ - \overline{\text{CI}}_- 
= \alpha_3 \left[ e^{-S^{(2)}(\rho_i)} - e^{-S^{(2)}(\rho_f)} \right],
\end{equation}
with $\alpha_3 \equiv \frac{d_A^2 - 1}{d_A^2 (d_A^2 + 1)} \frac{1}{d_\text{tot} (d_{\text{tot}}^2 - 1)} > 0$. Note that for $\delta t = 0^+$ there is no QCI between non-overlapping regions.
\end{enumerate}

We see that in both time regimes, the causal influence is stronger toward the direction with higher second R\'enyi entropy $S^{(2)}(\rho)$, providing quantitative evidence for the folklore that ``\textit{time flows toward increasing chaos},'' consistent with our results in Clifford circuits (Fig.~\ref{fig:light_cones}). In addition to the direction of the light cone, we also reproduce the exponential decay of QCI, $\sim \alpha_4 e^{-\delta t}$, with
\(
\alpha_4 = \frac{(d_A^2 - 1)(d_\text{tot}^2 d_A^{-2} - 1)}{d_A^2 (d_A^2 + 1) (d_\text{tot}^2 - 1)^3}
\left[ \tr \rho_f^2 - d_\text{tot}^{-1} \right]
\left[ (d_\text{tot}^2 + 1) - 2 d_\text{tot} \right] > 0 .
\)
See Fig.~\ref{fig:BGUE} for the full time dependence of $\overline{\text{CI}}$ in the BGUE model, as well as an illustration of $\Delta \text{CI}_{1,2,3}$ and the decay $\alpha_4 e^{-\delta t}$.


\subsection{Dual-unitary circuits}
\label{ssec:dual}

\begin{figure}
    \centering
    \includegraphics[]{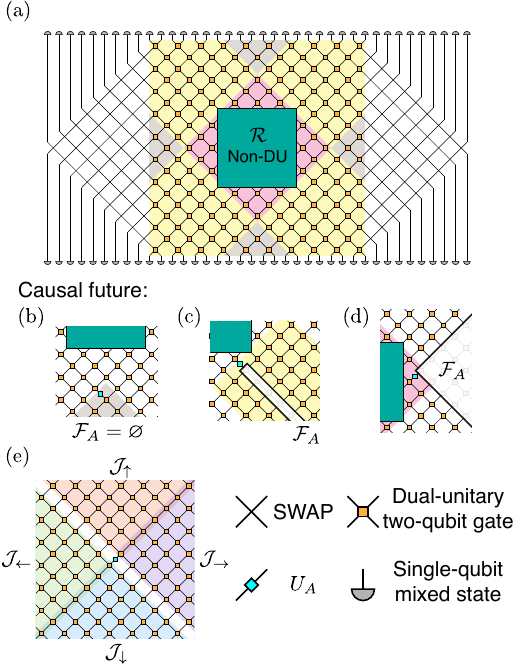}
    \caption{Causal structure in dual-unitary circuits. 
(a) Setup with maximally mixed states (or traced out) at both spatial and temporal boundaries, and a region with a prescribed state located in the middle of spacetime. The location of $A$ can be divided into three classes:
(b) the gray region has no locally detectable QCI;
(c) the yellow region has nonzero QCI only along a lightlike direction pointing radially outward;
(d) the pink region has nonzero QCI within a conic region pointing radially outward.
(e) Intuition for the causal structure: the perturbation $U_A$ can be canceled from any of the four conic regions. Hence, to obtain nonzero QCI, the ``non-DU'' region and the detection region $B$ must together block all four cones.}
    \label{fig:dual}
\end{figure}

Dual-unitary circuits are brickwork circuits composed of dual-unitary gates \cite{PhysRevLett.123.210601,Bertini_20201,Bertini_20202,PhysRevB.101.094304,bertini2025exactlysolvablemanybodydynamics}. Each two-qubit unitary gate has matrix elements $U_{ab}^{cd}$, where $a,b$ label the input qubit states and $c,d$ label the output qubit states, making $U_{ab}^{cd}$ a rank-four tensor. The dual-unitary condition requires that both $U_{ab}^{cd}$ and its reshaped tensor $\tilde{U}_{ac}^{bd} \equiv U_{ab}^{cd}$ be unitary matrices. Consequently, a dual-unitary circuit defines a unitary map along the time direction and, simultaneously, a unitary map along the spatial direction. This symmetry greatly simplifies the computation of QCI and allows us to analytically derive generic properties of the resulting causal structure.

In a dual-unitary circuit, the circuit itself does not single out a time direction, since both the vertical and horizontal directions can be regarded as time. As in previous sections, the arrow of time is set by the boundary conditions. We consider the setup shown in Fig.~\ref{fig:dual}(a), where the left and right boundaries of a dual-unitary circuit are connected to an environment via swap gates. In this setup, also used in Ref.~\cite{PhysRevX.12.011045}, the boundary conditions on all four sides of the circuit can be tuned through the choice of initial and final states. If the boundary conditions on all four sides are maximally mixed, a single-site unitary $U_A$ has no causal influence on any local single-site region $B$, because one can always conjugate $U_A$ by unitaries along a direction that does not include $B$, thereby moving $U_A$ into a multi-site unitary acting on a maximally mixed state. Nontrivial causal influence emerges only when the boundary conditions are nontrivial. Owing to the dual-unitary property of the gates, richer causal structures can arise. Rather than imposing conditions on one or more of the four boundaries, we instead consider a region in the middle of the circuit, illustrated by the green boxed region $\mathcal{R}$ in Fig.~\ref{fig:dual}(a). Within this region, we replace the gates with non-dual-unitary gates and include post-selections. Contracting the tensors in $\mathcal{R}$ then defines an effective boundary state at the edge of $\mathcal{R}$, enabling sites outside this region to exhibit nontrivial causal structure.


Below, we analyze the causal influence between pairs of single-site regions, denoted $A$ and $B$, in the configuration shown in Fig.~\ref{fig:dual}(a). We consider a generic fixed state in region $\mathcal{R}$ (which can be interpreted as the ``initial'' state in a radial picture), together with maximally mixed states on the outer boundary in all four directions. While the magnitude of the QCI depends on the specific ``initial'' state in $\mathcal{R}$, the shape of the causal future---that is, the region where QCI can be nonzero---can be determined independently of the details of the state. For each link $A$ in the circuit (see Fig.~\ref{fig:dual}(e)), we define four conic regions $\mathcal{J}_{\uparrow, \downarrow, \leftarrow, \rightarrow}$ separated by 45-degree lines. Note that these regions overlap along the light-ray directions. For right-moving links, $\mathcal{J}_\uparrow \cap \mathcal{J}_\rightarrow \neq \varnothing$ and $\mathcal{J}_\leftarrow \cap \mathcal{J}_\downarrow \neq \varnothing$, whereas for left-moving links, $\mathcal{J}_\uparrow \cap \mathcal{J}_\leftarrow \neq \varnothing$ and $\mathcal{J}_\rightarrow \cap \mathcal{J}_\downarrow \neq \varnothing$. The causal influence from $A$ to another link $B$ is determined by the relative positions of $B$ and $\mathcal{R}$ with respect to $A$. If any one of the four cones $\mathcal{J}_a$, with $a=\uparrow, \downarrow, \leftarrow, \rightarrow$, has no overlap with $\mathcal{R} \cup B$, then the causal influence vanishes, $\operatorname{CI}(A\!:\!B)=0$, since $U_A$ can be conjugated by gates within that cone and pushed to the outer boundary, where it is canceled. Equivalently, the ``future light cone of $A$,'' defined as the set of $B$ such that ${\rm CI}(A\!:\!B)\neq 0$, is given by
\begin{align}
    \mathcal{F}_A\equiv \left\{B:\left(\mathcal{R}\cup B\right)\cap \mathcal{J}_a\neq \varnothing,~
    \forall a=\uparrow, \downarrow, \leftarrow, \rightarrow \right\}
\end{align}

Based on this analysis, we can map out the future light cone $\mathcal{F}_A$ for different locations of $A$. There are three types of $A$, illustrated in gray, yellow, and pink in Fig.~\ref{fig:dual}(a). These three situations are depicted in Fig.~\ref{fig:dual}(b--d):
\begin{itemize}
    \item If $A$ is in the gray region, $\mathcal{R}$ blocks only one of the four conic regions. There is no future light cone, because it is impossible for any local $B$ to block the remaining three cones.
    \item If $A$ is in the yellow region, $\mathcal{R}$ blocks two of the four conic regions, so that $B$ must block the remaining two. Only links pointing away from $\mathcal{R}$ have a future light cone, which reduces to a single light ray (a 45-degree line).
    \item If $A$ is in the pink region, $\mathcal{R}$ blocks three of the four conic regions, so $B$ needs to block only the remaining one. Consequently, the entire remaining cone becomes the future light cone. In other words, a perturbation in the pink region has an ordinary future light cone, with the time direction pointing radially outward from $\mathcal{R}$.
\end{itemize}

The example of dual-unitary circuits illustrates that the time direction can not only be reversed but can also exhibit state-dependent changes in more than one dimension.

\section{Conclusion}
\label{sec:conclusion}

In this paper, we propose an experimentally measurable criterion for quantum causal influence and study it through simulations of Clifford circuits with final-state post-selection or in-circuit measurements. We also analyze analytically tractable models, including Brownian Gaussian unitary dynamics and dual-unitary circuits. Our results across multiple physical systems reveal the general principle that quantum causal structure is state-dependent, and that the arrow of time typically points from low-entropy to high-entropy regions. In the context of MIPT, we show that the range of QCI propagation probes the entanglement phase transition and peaks near the critical point. The results obtained in dual-unitary circuits further clarify that causal influence can be more general than simply pointing toward the future or the past.

Finally, we point out a few future directions worth exploring. 
\begin{enumerate}
\item {\bf Experimental realization.} Our proposal of XEQCI is sufficiently concrete to be realized on current quantum computing platforms. A particularly promising platform is superconducting qubits, which offer advantages in gate speed and programmability. Notably, similar cross-entropy--type experiments for MIPT have already been realized using superconducting qubits \cite{Hoke2023, PhysRevLett.134.120401}, and our proposed XEQCI experiment has essentially the same hardware requirements. Moreover, XEQCI experiments could also be implemented using neutral-atom arrays, which benefit from native mid-circuit measurements and highly reconfigurable geometries.

\item {\bf Other measures of QCI.} As discussed above, any state distinguishability measure that differentiates the superdensity operators $\varrho_B(U_A)$ can be used to quantify QCI. For example, for an ensemble of unitaries $U_A$, one may consider the Holevo information,
\begin{equation}
    \chi = S(\overline{\varrho_B}) - \Exp_{U_A} S\!\left( \varrho_B(U_A) \right),
\end{equation}
which quantifies the amount of classical information that can be transmitted from $A$ to $B$ by controlling $U_A$. It is of interest to study these alternative measures and assess whether any of them are suitable for experimental realization or other purposes. In the context of measuring QCI in monitored dynamics, it is also desirable to explore additional approaches for circumventing the post-selection problem. One possible approach is through amplitude-amplification quantum algorithms \cite{grover1996fast,wang2025postselection}.

\item {\bf Emergent unitarity.} As seen in the case of monitored quantum circuits, when the dynamics is non-unitary, QCI is generally nonzero between all pairs of points, rendering the future light cone ill-defined. Causal influence is a particular form of correlation, corresponding to the case where the operation in $A$ is unitary. If one inserts a pair of generic operators in regions $A$ and $B$, one simply measures correlations between these two regions. The distinction between causal influence and correlation is well defined only because the underlying dynamics is unitary: inserting $U_A$ into the tensor network without any other operators leaves the value of the network invariant. Only in this case can one then insert another operator $O_B$ and meaningfully discuss the influence of $U_A$ on $O_B$. An interesting question is whether the notion of causal influence can be generalized to systems in which unitarity is emergent. For example, in discussions of black hole final states \cite{horowitz2004black}, the presence of a final-state projection renders the dynamics non-unitary, yet unitarity approximately holds for local operators. This situation is reminiscent of our result in Fig.~\ref{fig:light_cones}(g), where despite the absence of exact unitarity, an approximately one-sided light cone emerges due to scrambling of the future state.

\item {\bf Nonlocality of QCI.} Two key features of QCI, compared to classical causal influence, are state dependence and nonlocality. In this paper, we have mainly focused on causal influence between pairs of local regions. The nonlocality of QCI has been explored in Refs.~\cite{Cotler_2019,yates2025localarrowstimequantum}, including examples such as quantum teleportation. It would be interesting to further investigate the properties of QCI for nonlocal regions and to understand its relation to other concepts, such as quantum error correction and emergent spatial locality in holographic duality.

\item {\bf Relation with holographic duality.} Based on concrete models such as random tensor networks \cite{hayden2016holographic,qi2018space}, it would be interesting to understand the relation between QCI and emergent causal structure in quantum gravity theories, especially in geometries such as de Sitter space and black hole interiors, where a systematic understanding of holographic duality remains an open problem.
\end{enumerate}


\acknowledgements

We thank Sarang Gopalakrishnan, Vedika Khemani, Yaodong Li and Yi-Zhuang You for helpful discussion. 
The stabilizer simulation within is based on the Julia package \texttt{QuantumClifford}. H.-Y. W. is supported by NSF QuSEC-TAQS OSI 2326767. H.T. is supported by Shoucheng Zhang Graduate Fellowship Program. X.-L. Q. is supported by Simons Foundation.

\appendix



\section{Numerical recipe on stabilizer simulation}
\label{app:stabilizer}



\begin{figure}[ht]
    \centering
    \includegraphics{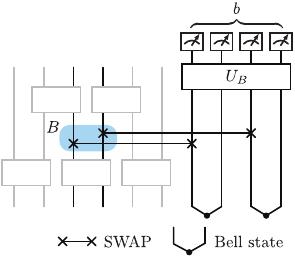}
    \caption{Convention for the POVM in region $B$ used to detect QCI. }
    \label{fig:POVMinB}
\end{figure}

In this section, we elaborate on the numerical procedure for simulating the XEQCI experiment using stabilizers. Our goal is to compute the XEQCI averaged over a class of evolutions (e.g., random Clifford circuits, CSS circuits, or hybrid circuits), i.e.,
\begin{equation}    \label{eq:chibar_repeat}
    \overline{\chi} = \Exp_{\mathcal{C}} \sum_{b,m} \Exp_{U_A} \Pr(b,m,U_A)\, c_{b,m} , 
\end{equation}
where $\mathcal{C}$ denotes a class of quantum circuits together with the initial and final states over which the average is taken, and $c_{b,m}$ is the classically simulated quantity defined in Eq.~\eqref{eq:csim}. It follows that, in a simulation of this experiment, we require a procedure to sample $b$ and $m$ according to the probability distribution $\Exp_{U_A}\Pr(b,m,U_A)$ (quantum simulation), as well as to compute $c_{b,m}$ (classical simulation). In addition, we must specify the POVM used in region $B$ to obtain the outcomes $b$. To ensure that the POVM is sufficiently comprehensive, we adopt the convention illustrated in Fig.~\ref{fig:POVMinB}.

A verbal description of Fig.~\ref{fig:POVMinB} is given below. Let the time coordinate of region $B$ be $t_B$, and let its spatial extent be $L_B$. In addition to the evolving system, we prepare $L_B$ Bell pairs on $2L_B$ auxiliary qubits. At time $t_B$, we swap region $B$ with one half of each Bell pair. We then apply a random Clifford gate acting on the $2L_B$ qubits, followed by measurements of each of the $2L_B$ qubits in the $Z$ basis. The random Clifford gate is introduced to ensure the generality of QCI detection; without it, this POVM would be sensitive only to effects that influence the computational basis. 

The approach for the quantum simulation is:
\begin{enumerate}
    \item Sample a circuit realization.
    \item Simulate the evolution up to the final time. If there are any measurements in the circuit besides the measurement in region $B$, we record their outcomes collectively as $m$. At times $t_A$ and $t_B$, perform the following operations:
    \item At time $t_A$, apply a completely depolarizing channel in region $A$.
    \item At time $t_B$, apply the measurement shown in Fig.~\ref{fig:POVMinB} to obtain outcome $b$.
\end{enumerate}
The use of a completely depolarizing channel in region $A$ is justified as follows. Equation~\eqref{eq:chibar_repeat} is linear in $\Pr(b,m,U_A)$, which is in turn linear in $U_A \otimes U_A^\dagger$. We therefore average over random $U_A$ forming a unitary 1-design. Averaging $U_A \otimes U_A^\dagger$ over a 1-design yields a completely depolarizing channel.

To compute the classical quantity $c_{b,m}$, we use the following useful property of stabilizer states: for any computational-basis measurement, the Born probabilities of all possible outcomes take a uniform value $1/2^n$ for some integer $n$. As a result, $c_{b,m}$ can take only the values $0$ or $1$. This property also ensures that the standard deviation of $c_{b,m}$ is bounded by $1$, so the experiment can be repeated to reduce the statistical uncertainty of $\overline{\chi}$, which scales as $1/\sqrt{N}$ by the central limit theorem, where $N$ is the number of sampled values of $c_{b,m}$. Therefore, the numerical procedure is as follows:
\begin{enumerate}
    \item Classically simulate the circuit chosen in the quantum simulation.
    \item If there are any measurements of type $m$ besides the detector in region $B$, post-select the measurement outcome to be $m$. If the Born probability of outcome $m$ is zero, discard the current trial and restart. At time $t_B$, perform the operation described in step~3.
    \item Apply the SWAP gates and $U_B$ as described in Fig.~\ref{fig:POVMinB}, and post-select the measurement outcome to be $b$. If post-selection succeeds, set $c_{b,m}=1$; if it fails, set $c_{b,m}=0$.
\end{enumerate}

\section{The CSS gate set}
\label{app:CSS}

All Clifford gates can be specified by their action on stabilizers. CSS gates are defined by the property that $X$ stabilizers are mapped to $X$ stabilizers and $Z$ stabilizers are mapped to $Z$ stabilizers. In Fig.~\ref{fig:light_cones}(d,e,f), we study brickwork circuits composed of random two-qubit CSS gates. Here, we elaborate on how these gates are sampled.

The set of all two-qubit CSS gates is relatively small, and we enumerate it below. It consists of the six basic gates listed in Table~\ref{tab:CSS}. Additionally, each entry can have an independent $\pm$ sign, and the resulting Clifford gate is still CSS. As a result, the total number of two-qubit CSS gates is $6 \times 2^4 = 96$. In the simulations shown in Fig.~\ref{fig:light_cones}(d,e,f), we sample these gates uniformly.

\begin{table}[h!]
\centering
\begin{tabular}{l|cccc}
\hline
Gate & $X_1$       & $X_2$         & $Z_1$        & $Z_2$         \\ 
\hline
$g_0$ ($\id$)              & $X_1$       & $X_2$       & $Z_1$       & $Z_2$        \\
$g_1$ (SWAP)           & $X_2$       & $X_1$       & $Z_2$       & $Z_1$        \\
$g_2$ ($\mathrm{CNOT}_{1\to 2}$)           & $X_1 X_2$    & $X_2$       & $Z_1$       & $Z_1 Z_2$     \\
$g_3$ ($\mathrm{CNOT}_{2\to 1}$)   & $X_1$       & $X_1 X_2$    & $Z_1 Z_2$    & $Z_2$        \\
$g_4$                   & $X_2$       & $X_1 X_2$    & $Z_1Z_2$       & $Z_1$        \\
$g_5$                   & $X_1 X_2$    & $X_1$       & $Z_2$       & $Z_1 Z_2$     \\
\hline
\end{tabular}
\caption{Stabilizer-mapping rules for two-qubit CSS gates. }
\label{tab:CSS}
\end{table}

\section{QCI landscape data for fitting correlation time in MIPT}
\label{app:MIPTfullA}

In this appendix, we provide the QCI landscape data used to extract the correlation time $\tau$ shown in Fig.~\ref{fig:correlation_time}.

First, we restate the circuit setup. We simulate a brickwork circuit of random Clifford gates. At each site and at each time step, a computational-basis measurement is inserted with probability $p$.

The correlation-time data are obtained as follows. We take the influence source $A$ to be the entire Hilbert space at time $t_A = T/2$, where $T$ is the total simulation time, so that the resulting QCI landscape is translationally invariant in space and thereby admits a well-defined correlation time. We fix the system size to $L = 60$ and vary $T$ and $p$. We then average $\chi$ over the spatial coordinate to obtain $\overline{\chi}$ as a function of time $t$. The data are fit to
\(
1 - \overline{\chi}(t) = c\, e^{-|t - t_A|/\tau},
\)
from which the correlation time $\tau$ is extracted.

\begin{figure*}[t]
    \centering
    \includegraphics[]{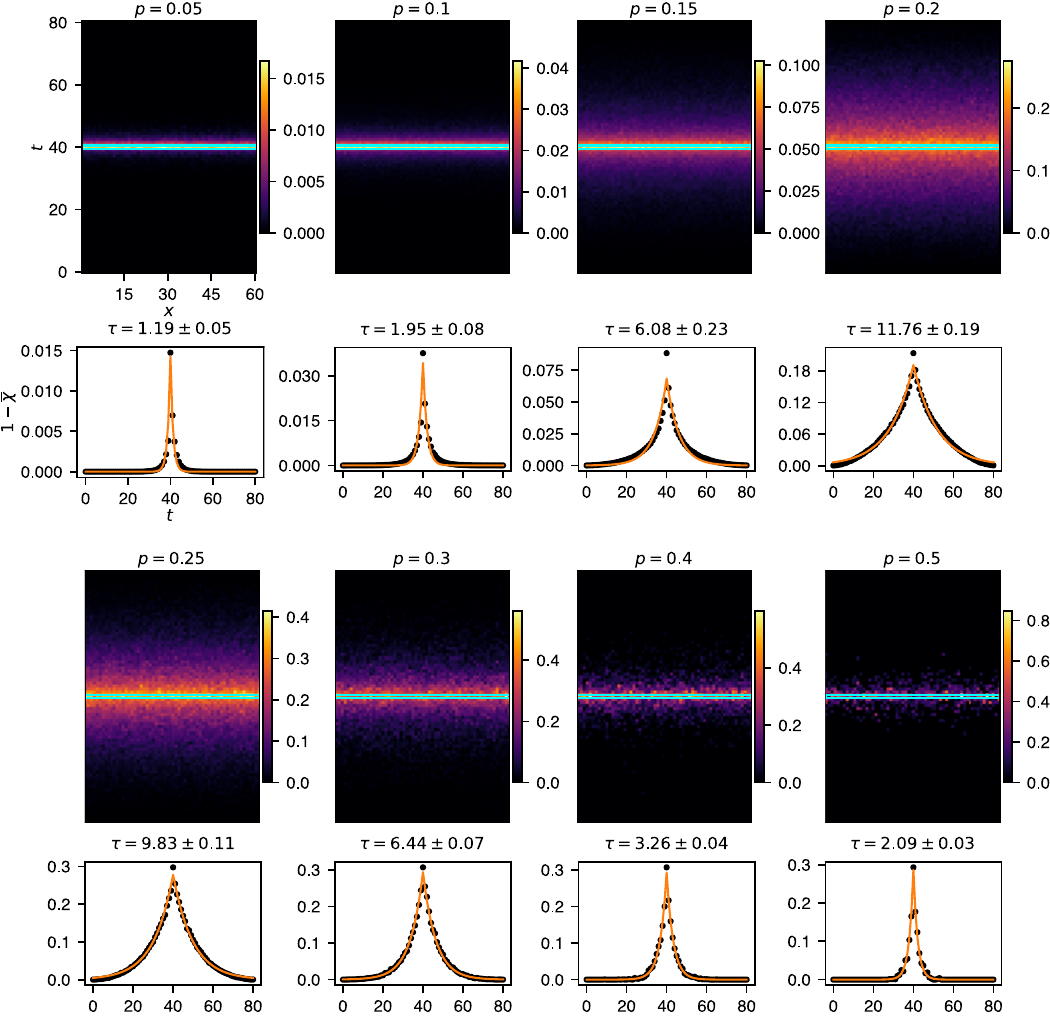}
    \caption{QCI landscape and correlation time fitting in hybrid circuits at different measurement rate $p$. The number of qubits is $L=60$, and the total simulation time is $T=80$. }
    \label{fig:MIPT_landscape_fit}
\end{figure*}

Representative data obtained using this procedure at $L = 60$ and $T = 80$ are shown in Fig.~\ref{fig:MIPT_landscape_fit}.

\section{Calculation of QCI in Brownian Gaussian unitary ensemble}
\label{app:BGUE}

Here, we present the detailed calculation of the averaged QCI in the Brownian Gaussian unitary ensemble (BGUE) toy model, using results from Ref.~\cite{Tang2024kpv}.

First, we restate the setup from the main text. Specifically, we consider an initial state $\rho_i$ evolved under BGUE for a time $T$, and then post-selected onto a final state described by the density matrix $\rho_f$. We insert a Haar-random unitary $U_A$ in region $A$ (with dimension $d_A$) at time $t_U$ ($0 < t_U < T$), and a random Hermitian operator $O_B$ in region $B$ (with dimension $d_B$) at time $t_O \equiv t_U \pm \delta t$ ($\delta t > 0$). Recall that $\mathcal U_t \equiv \mathcal P e^{-i \int_0^{t} dt' \, H(t')}$ is the time-ordered exponential; the corresponding superdensity operator is then explicitly given by:
\begin{equation}
\varrho(U_A)(O_B,O_B^\dagger)=\operatorname{Tr}(\rho_f\mathcal W\rho_i\mathcal W^\dagger)
\end{equation}
\begin{equation}
\mathcal W:=\begin{cases}
\mathcal U_{T-t_O}O_B\mathcal U_{t_O-t_U}U_A\mathcal U_{t_U},\ t_O>t_U,\\
\mathcal U_{T-t_U}U_A\mathcal U_{t_U-t_O}O_B\mathcal U_{t_O},\ t_U>t_O
\end{cases}
\end{equation}
Based on Eq.~\eqref{eq:cotler_avg}, we define the averaged CI by further averaging over the BGUE evolution:
\begin{equation}   
    \begin{aligned}
    \overline{\mathrm{CI}}(A:B) &:= \mathbb E_{\mathcal U\sim \text{BGUE}}\int dU_A \int dO_B \\
    &\left| \varrho(U_A)(O_B, O_B^\dagger) - \int d{U_A} \, \varrho(U_A)(O_B, O_B^\dagger) \right|^2
    \end{aligned}
\end{equation}
Since the integrand only contains two copies of $\mathcal U_t$ and $\mathcal U_t^\dagger$, the average of such variables over the BGUE has been worked out explicitly in Ref.~\cite{Tang2024kpv} :
\begin{equation}
\mathbb E_{\mathcal U\sim \text{BGUE}}[\mathcal U_t^{\otimes 2}\otimes \mathcal U_t^{*\otimes 2}]\equiv \sum_{i=1}^{8}f_i(t)A_i
\end{equation}
where $A_i$ are a particular set of permutation operators on four copies of Hilbert space and $f_i(t)$ are corresponding time-dependent coefficients (see Eqs.~(2.1), (2.10), (2.14) and (2.15) of Ref.~\cite{Tang2024kpv} for explicit definition). The remaining averages over $U_A,O_B$ also give rise to certain permutation operators~\cite{Cotler_2019}. Finally, we compute the linear superposition of the trace of the product of permutation operators and obtain $\overline{\text{CI}}(A:B)$, which is implemented in an automated code using Mathematica.

\newpage
\bibliography{references}

\end{document}